# MERLIN-SUITE: Probabilistic modular GRN inference from multi-omics data integrating regulatory priors and transcription factor activity

**Suvojit Hazra[1], Marina Kotvanova[1], Kirstan Gimse[1], Sushmita Roy[1,2]**

[1]Wisconsin Institute for Discovery, University of Wisconsin-Madison, Madison, WI, USA
[2]Department of Biostatistics and Medical Informatics, University of Wisconsin-Madison, Madison, WI, USA

## Abstract

Accurately reconstructing gene regulatory networks (GRNs) is essential for understanding transcriptional processes in development and disease. MERLIN-SUITE (https://github.com/Roy-lab/MERLIN-SUITE) represents a collection of algorithmic extensions based on MERLIN (Modular regulatory network learning with per gene information) a probabilistic framework that infers gene-specific and module-specific regulatory programs of co-regulated modules, capturing both detailed and modular aspects of transcriptional networks. While expression-based inference is effective, it often aligns poorly with experimentally validated regulatory interactions. MERLIN-P addresses this by integrating external regulatory priors, such as motif, ChIP, and perturbation data, to enhance biological relevance and predictive accuracy. MERLIN-P-TFA further advances the framework by incorporating regularized estimation of latent transcription factor activity (TFA), overcoming the limitation that TF mRNA levels may not represent protein activity. By integrating expression data, prior knowledge, and activity-aware modeling, this unified approach supports robust GRN reconstruction in both bulk and single-cell datasets. This chapter presents the MERLIN-SUITE with a focus on MERLIN-P-TFA and demonstrates its use on a single-cell, multi-modal dataset of mouse cellular reprogramming to infer GRNs and identify key regulators.

# 1. Introduction

Gene expression is regulated through coordinated interactions between transcription factors and their binding sites in gene promoters and enhancers, as well as among transcription factors (TFs), signaling proteins and chromatin remodelers [1, 2]. These interactions shape transcriptional programs that control cellular identity, development, and responses to environmental or stress stimuli, which can be altered in disease or modulated by therapy [3]. Inferring Gene Regulatory Networks (GRNs) from large-scale omics data provides a systems-level framework to model these complex regulatory interactions, representing regulators and their target genes as nodes connected by directed edges [1]. Traditionally, GRN inference models have relied solely on RNA expression data (bulk or single cell), which often fail to recover experimentally validated interactions, partly because structure learning is a heuristic search and partly because TF mRNA abundance may not reliably reflect protein activity [4].

To address these challenges, integrative approaches that combine gene expression with additional regulatory evidence have become increasingly important. These approaches include incorporating regulatory prior information such as motif occurrence, ChIP-seq binding data, and perturbation studies, as well as modeling latent transcription factor activity (TFA), especially useful when TF transcript levels and its activity are not correlated due to post-translational effects [5, 6]. MERLIN (Modular regulatory network learning with per gene information) provides a probabilistic framework that infers gene-specific regulatory programs while constraining genes into co-regulated modules, capturing both fine-grained and modular organization of transcriptional networks [7]. MERLIN-P extends this framework by integrating external regulatory priors [8], and MERLIN-P-TFA further enhances it by incorporating regularized estimation of latent TF activity [9]. Together, these advances enable robust reconstruction of GRNs from bulk [10] and single-cell [6] multi-omics datasets.

In this book chapter, we present MERLIN-SUITE (https://github.com/Roy-lab/MERLIN-SUITE), which includes MERLIN, MERLIN-P, MERLIN-P-TFA, and additional utilities, EstimateNCA and MERLIN-VIZ for TFA estimation and interpretation. We provide the theoretical foundation of the MERLIN framework and demonstrate the application of MERLIN-P-TFA using an existing single-cell, multimodal dataset of mouse cellular reprogramming [11] (**Figure 1**). We also present four types of network visualizations (zeromean expression-based, cell-type-specific expression-based, condition-specific expression-based, pseudobulk expression-based) for interpreting MERLIN-inferred networks, including the R Shiny-based MERLIN-VIZ application (https://github.com/Roy-lab/MERLIN-VIZ). This step-by-step guide is designed to help users apply MERLIN-P-TFA to other systems for modeling modular gene regulatory networks (GRNs). Earlier version of the framework, MERLIN (https://github.com/Roy-lab/merlin) [7], is also available for users who wish to apply implementations based solely on gene expression profiles without incorporating regulatory prior information.

# 2. The theory of MERLIN

MERLIN employs a probabilistic graphical model to infer regulatory network by maximizing the likelihood of observed expression data given learned network structure and module assignments [7]. Let $X = \{X_1, \ldots, X_n\}$ denote random variables representing gene expression levels and let $\mathcal{R} \subseteq \mathcal{X}$ be the subset designated as candidate regulators. MERLIN jointly infers (i) the graph structure $\mathcal{G}$ describing the regulatory relationships between genes and regulators, (ii) module membership $M$ of each gene, and (iii) parameters $\Theta$ that best explain the given dataset $\mathcal{D}$. The posterior over these quantities is proportional to the data likelihood $P(\mathcal{D}|\mathcal{G}, \Theta, M)$ and chosen priors – prior distribution of parameters $P(\Theta|\mathcal{G}, M)$ and a prior distribution over the graph $P(\mathcal{G}|M)$ [$P(M)$ is assumed to be a uniform prior not influencing the total probability]:

$$P(\mathcal{G}, M, \Theta|\mathcal{D}) \propto P(\mathcal{D}|\mathcal{G}, \Theta, M)P(\Theta|\mathcal{G}, M)P(\mathcal{G}|M)P(M)$$

The graph prior $P(\mathcal{G}|M)$ factorizes over potential regulator-target edges, combining contributions from edges present in $\mathcal{G}$ and those absent:

$$P(\mathcal{G}|M) = \prod_{X_j \to X_i \in \mathcal{G}} P(X_j \to X_i|M_i) \prod_{X_j \to X_i \notin \mathcal{G}} 1 - P(X_j \to X_i|M_i)$$

where $X_i$ denotes a target gene and $X_j \in \mathcal{R}$ corresponds to a regulator.

MERLIN models the prior probability of an edge using a logistic form:

$$P(X_j \to X_i|M_i) = \frac{1}{1 + \exp\left(-\left(p + \beta^R f_{ij} + \sum_k \beta^k \times w_{ij}^k\right)\right)}$$

Here, the sparsity parameter $p$ controls how strongly dense graphs are penalized (more negative values encourage sparser networks), $\beta^R$ controls the effect of the module prior and $f_{ij}$ measures the extent to which regulator $X_j$ regulates other genes in the $X_i$'s module $M_i$. $f_{ij} = \frac{d_{jM_i}}{d_j}$ where $d_{jM_i}$ is the number of predicted targets of $X_j$ in the $X_i$'s module and $d_j$ is the total number of predicted targets of $X_j$. When integrating additional prior networks, each prior can contribute via its importance term $\beta^k$ and weight/confidence term $w_{ij}^k$ which controls the confidence in the edge in the prior network, so stronger prior evidence increases the prior probability of including the edge $X_j \to X_i$ in predicted network.

Because the underlying structure is a dependency network, MERLIN uses a pseudo-likelihood approximation to the full likelihood [12] to model the data likelihood, written as a product of per-gene conditional distributions for each random variable $X_i$ given its parents $R_i$ estimated as multivariate conditional Gaussians:

$$P(\mathcal{D}|\mathcal{G}, \Theta, M) = \prod_{i=1}^{n} \prod_{d=1}^{|\mathcal{D}|} P(X_i = x_i^d | R_i = x_{R_i}^d, \theta_i)$$

where $n$ is the number of all (target) genes, $\theta_i$ are the parameters associated with conditional distributions $P(X_i|R_i)$ and $x_{R_i}^d$ are expression values of $R_i$ in sample $d$.

The learning procedure begins from an initial module assignment (often seeded from expression-based clustering) and then iterates until convergence between two updates: (1) updating the regulator set for each gene given the current module assignments, adding only edges that improve the overall score of the network; and (2) revising module memberships using both expression similarity and the currently inferred regulatory structure. If the user does not provide an initial module file, MERLIN randomly partitions genes to initialize module assignments.

In MERLIN-P settings [8], regulator selection can be augmented by estimates of latent transcription factor activity (TFA) for regulators present in the input prior network and combining those with data $\mathcal{D}$. To estimate TFA, MERLIN framework uses variant of network component analysis (NCA) [9, 13] which uses matrix decomposition, and optimizes the following objective:

$$min(\ ||E_{n\times|\mathcal{D}|} - A_{n\times m} P_{m\times|\mathcal{D}|}\ ||_2^2 + \lambda\ ||\ A_{n\times m}(W_{n\times m})^T ||_1)$$

where $E$ is the given expression matrix, $A$ is the adjacency matrix encoding the prior network structure, $n$ is the number of target genes, $m$ is the number of regulators, $P$ is the TFA matrix for the regulators, $W_{ij}$ penalizes edges with weak prior support, and $\lambda$ controls the regularization strength. When $\lambda$= 0, the objective reduces to the unregularized NCA approach [13]. When $\lambda$≠0, the objective becomes the regularized NCA-LASSO approach [9], which encourages sparse and biologically plausible regulatory interactions. Optimizing the objective yields both the TFA matrix $P$, which represents inferred TFA (with rows corresponding to regulators, columns to samples or cells, and elements to inferred activity scores), and an updated regulatory weight matrix $A$ that quantifies the strength of regulator-target gene interactions.

The MERLIN-P-TFA package [9] integrates two components: (i) unregularized [13] or regularized [9] EstimateNCA (https://github.com/Roy-lab/EstimateNCA) and MERLIN-P (https://github.com/Roy-lab/merlin-p) [8] packages. In the MERLIN-P-TFA framework, the estimated TFA matrix $P$ is concatenated with the original expression matrix $E$ to form an augmented expression matrix. In addition, the regulatory prior network is expanded to include both the original regulators and the inferred TFAs. The resulting augmented expression matrix and updated regulatory prior network are then used as inputs to the MERLIN-P framework [8] for the inference of context-specific modular gene regulatory networks (**Figure 1**).

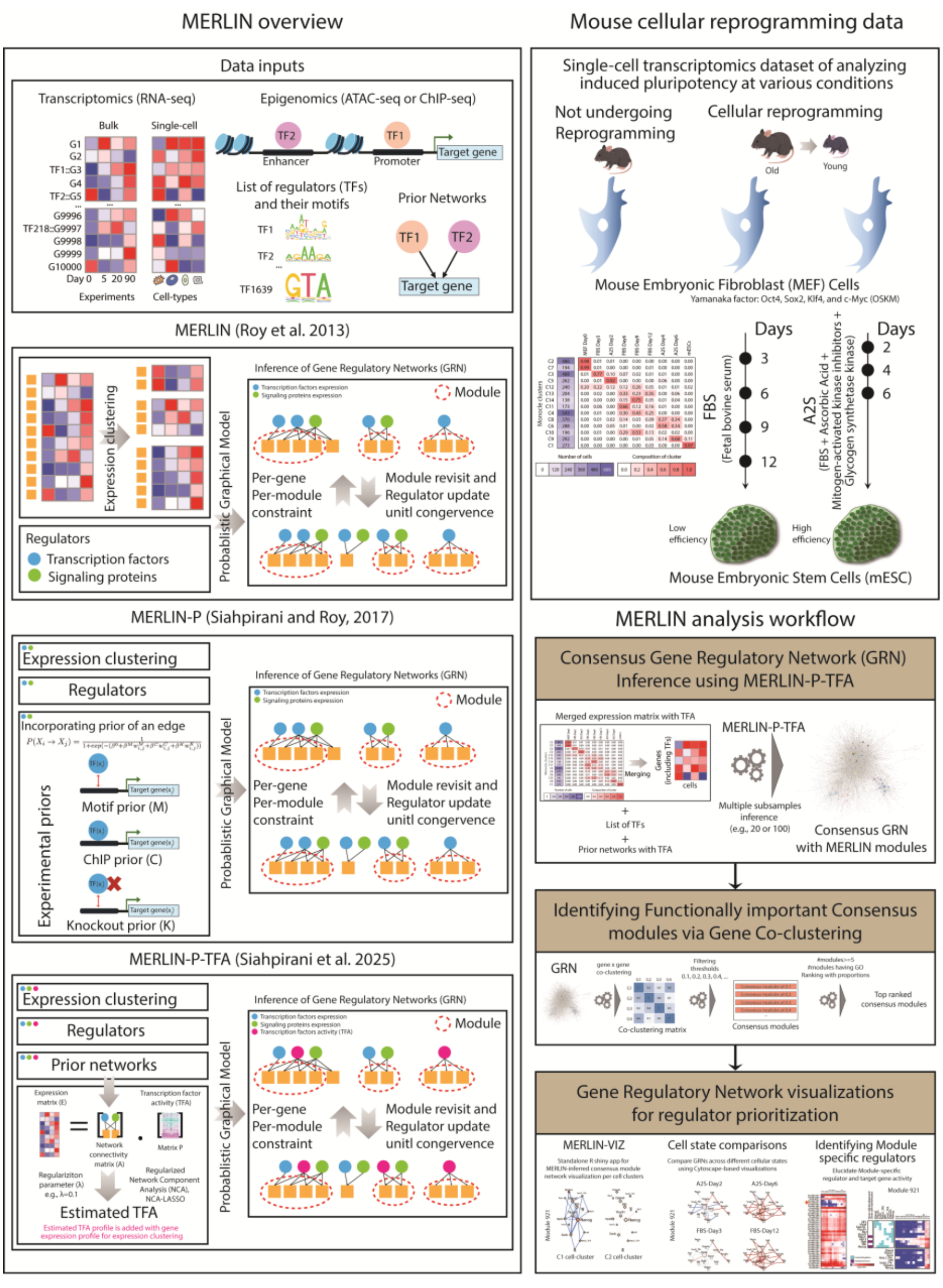

*Figure 1: Overview of the MERLIN framework for probabilistic modular gene regulatory network (GRN) inference integrating expression, regulatory priors, and transcription factor activity (TFA),*

***and application of MERLIN-P-TFA to mouse cellular reprogramming. (a)*** *The overview illustrates the evolution and integration of MERLIN, MERLIN-P, and MERLIN-P-TFA within a unified framework for context-specific modular GRN reconstruction. Starting from transcriptomic data (bulk or single-cell RNA-seq), MERLIN applies a probabilistic graphical model that jointly infers gene-specific regulatory programs while constraining genes within co-regulated modules through iterative module refinement and regulator updates. MERLIN-P extends this framework by incorporating external experimental regulatory priors, including motif, ChIP-seq, and perturbation evidence, to improve the biological consistency of inferred edges. MERLIN-P-TFA further enhances inference by estimating latent transcription factor activity (TFA) using prior-guided unregularized or regularized Network Component Analysis (NCA or NCA-LASSO) and integrates TFA profiles with gene expression for activity-aware clustering and network learning.* ***(b)*** *MERLIN-P-TFA was applied to clustered single-cell RNA sequencing data from mouse reprogramming fibroblasts (MEFs) [11]. The dataset consists of nine samples from different timepoints and conditions: MEFs not undergoing reprogramming (MEF Day0); MEFs reprogramming in fetal bovine serum (FBS) collected on Days 3, 6, 9, and 12 (FBS Day3, FBS Day6, FBS Day9, FBS Day12); MEFs reprogramming in FBS with ascorbic acid, mitogen-activated kinase inhibitors, and glycogen synthetase kinase (A2S) collected on Days 2, 4, and 6 (A2S Day2, A2S Day4, A2S Day6); and mouse embryonic stem cells (mESCs). These data were clustered into 14 cell clusters (C1–C14) using Monocle.* ***(c)*** *The analysis workflow on the mouse reprogramming includes (i) consider inputs the dataset and with multiple subsampling iterations (20 times in the present study) generating the consensus GRNs, (ii) functionally (GO) enriched large consensus modules (>=5) based on gene co-clustering on the consensus GRN and their ranking to obtain top modules for downstream analysis, (iii) network visualization for celltype-specific manner using MERLIN-VIZ, sample-type specific manner using Cytoscape and Module specific regulators from pseudobulk expression profile of modules per cell-cluster.*

## 3. Materials

### *3.1. MERLIN-SUITE software package*

All the three version of MERLIN (MERLIN [7], MERLIN-P [8], MERLIN-P-TFA [9]) and its related utility EstimateNCA [9, 13], are part of the MERLIN-SUITE package. These tools are developed in C++, compiled and tested for Linux environment. The source codes of MERLIN-SUITE is freely available for academic use in Roy Lab GitHub repository (https://github.com/Roy-lab/MERLIN-SUITE). Compilation requires the GNU Compiler Collection (GCC ≥ v6.3.1) and the GNU Scientific Library (GSL ≥ v2.6). The current implementation of MERLIN has been tested with GSL version 2.6, and the MERLIN-SUITE repository includes a compatible bundled GSL library. Users may either install a compatible system version of GSL (e.g., GSL v2.6) or use the bundled library included with MERLIN-SUITE. When using the bundled library, the library path should be added before running the executables:

```
export LD_LIBRARY_PATH=${LD_LIBRARY_PATH}:<path_to_merlin>/gsl_lib
```

On Ubuntu/Debian systems, GSL can be installed using:

```
sudo apt install libgsl-dev
```

MERLIN-SUITE supports multiple input configurations depending on the version. MERLIN [7] is purely expression-based and requires: (i) a normalized bulk or single-cell RNA-seq gene expression matrix, (ii) a regulator list consisting of a subset of genes from the expression matrix , and (iii) an initial module assignment of constituent genes from the expression matrix. MERLIN-P [8] additionally requires (iv) a regulatory prior network, which can be derived from bulk or single-cell chromatin accessibility data (e.g., ATAC-seq, ChIP-seq, or perturbation studies) via transcription factor (TF) motif enrichment. In the absence of chromatin accessibility data, MERLIN-P [8] can still be applied using RNA-seq data together with species-specific regulator lists based on sequence-specific TF motifs. MERLIN-P-TFA extends this framework by incorporating inferred TF activities as additional regulatory features.

### *3.2. Mouse cellular reprogramming dataset*

In the present study, MERLIN-P-TFA is demonstrated on a mouse cellular reprogramming dataset [11] (**Figure 1**). This dataset comprises single-cell transcriptomic profiles from nine (i-ix) experimental conditions representing distinct stages and reprogramming efficiencies. These include: (i) mouse embryonic fibroblasts not undergoing reprogramming (MEF Day 0); (ii-v) MEFs reprogrammed under fetal bovine serum (FBS) conditions with low reprogramming efficiency, collected at Days 3, 6, 9, and 12 (FBS Day 3, Day 6, Day 9, and Day 12); (vi-viii) MEFs reprogrammed under accelerated, high-efficiency conditions using FBS supplemented with ascorbic acid, MEK inhibitors, and GSK3 inhibitors (A2S), collected at Days 2, 4, and 6 (A2S Day 2, Day 4, and Day 6); and (ix) fully reprogrammed mouse embryonic stem cells (mESCs).

Cells from all conditions were jointly analyzed and clustered into 14 transcriptionally distinct cell clusters (C1-C14) using Monocle as described in **Figure 1** [11]. The dataset is provided to MERLIN-P-TFA in three input components: (i) a merged expression matrix of 2,231 variables (2,100 genes and 131 TFA regulators) across 4,633 cells spanning the fourteen cell types (C1-C14) (see **Section 4.4.1**); (ii) a regulator list of mouse (*Mus musculus*) comprising 2,814 factors (2,683 original mouse regulators and 131 inferred TFA regulators) (see **Section 4.4.3**); and (iii) a regulatory prior network containing 8,870,126 edges (see **Section 4.4.4**).

Each MERLIN-P-TFA run (see, **Section 4.4.6**) was performed on a standard desktop computer with ≥1 GB of available memory and required approximately 1 hour under typical parameter settings.

## 4. Methods

In this book chapter, we illustrate the full MERLIN-SUITE workflow using the MERLIN-P-TFA framework, which integrates MERLIN-P, and estimateNCA components into a unified analysis pipeline. To perform MERLIN-P-TFA, The methodological workflow of MERLIN-SUITE on the mouse cellular reprogramming single-cell dataset consists of six main steps: (1) downloading and installing EstimateNCA and MERLIN-P (Section **4.1**), (2) preparing input files for EstimateNCA (Section **4.2**), (3) applying EstimateNCA to obtain the estimated transcription factor activity (TFA) profiles (Section **4.3**), (4) preparing input files for MERLIN-P-TFA (Section **4.4**), (5) applying MERLIN-P using the estimated TFA profiles for MERLIN-P-TFA network inference (Section **4.5**), and (6) evaluating and validating the inferred networks using MERLIN-VIZ and other visualization protocols (Section **4.6**). These steps are demonstrated on the mouse cellular reprogramming dataset.

### *4.1. Downloading and installing EstimateNCA and MERLIN-P*

The current implementation of MERLIN-P-TFA is supported on Unix-based operating systems. The following Unix commands can be used to download and install EstimateNCA and MERLIN-P via the command-line terminal.

```
#Downloading and installing EstimateNCA application
git clone https://github.com/Roy-lab/EstimateNCA.git
cd EstimateNCA/
make
#Verify installation
./NCALearner
#Downloading and installing MERLIN-P application
```

```
git clone https://github.com/Roy-lab/merlin-p.git
cd merlin-p/
make
#Verify installation
./merlin
```

After compilation is complete, the main EstimateNCA executable will be available at `EstimateNCA/NCALearner`, and the MERLIN-P executable will be available at `merlin-p/merlin`.

### *4.2. Preparing input files for EstimateNCA*

EstimateNCA requires a gene expression matrix, a list of regulators, a list of target genes, and a prior regulatory network to infer latent transcription factors activities (TFAs) of regulatory transcription factors (see **Theory section** for detailed methodology). For mammalian cell lines, the EstimateNCA application is usually performed using 100 random initializations (`Rand_init = 100; e.g., Rand_init_0- Rand_init99`). EstimateNCA requires the input files in the following format:

#### *4.2.1. Expression file*

The example expression file (`expression.txt`) used throughout this protocol is provided in the MERLIN-SUITE GitHub repository (https://github.com/Roy-lab/MERLIN-SUITE) together with the remaining example input files. Users may directly use this file to reproduce the example workflow presented in this chapter or replace it with their own normalized expression matrix. The input expression file (`expression.txt`) should be provided as a tab-delimited gene-by-cell expression matrix without a header (cell metadata is not required). For single-cell expression data, the matrix should be merged across all cell clusters. For bulk expression data, the cell dimension should instead represent experimental conditions or biological samples.

```
Sept11	2.184866	3.061474	2.237097	1.874197 …
Sep15	2.983654	4.238418	3.770023	3.214221 …
Marc2	1.077569	1.325657	0.894839	0.909119 …
Sept7	2.711823	2.964259	2.324860	2.874887 …
Aars	2.094418	2.209428	1.095949	1.113439 …
…
Zic3	0.000000	0.000000	0.000000	0.000000 …
Zscan10	0.000000	0.000000	0.000000	0.000000 …
Zwilch	0.718379	1.325657	1.000460	0.000000 …
Zwint	1.015942	2.072627	1.095949	1.437443 …
Zyx	2.983654	4.592210	2.899609	4.430498 …
```

#### *4.2.2. List of regulators file*

The list of regulators file (`regulators.txt`) should contain a single column specifying the regulator names, with one regulator per line.

```
0610010K14Rik
100041979
1700024P04Rik
1700054O13Rik
1810007M14Rik
…
Zscan21
```

```
Zscan22
Zscan4c
Zswim4
Zzz3
```

### 4.2.3. List of target genes file

Similar to the regulator list, the target gene list file (`targets.txt`) should contain a single column specifying target gene names, with one target gene per line.

```
Sept11
Sep15
Marc2
Sept7
Aars
…
Zic3
Zscan10
Zwilch
Zwint
Zyx
```

### 4.2.4. Prior Network file

The prior regulatory network should be cell-type agnostic and can be derived from bulk or single-cell ATAC-seq, ChIP-seq, perturbation assays, or sequence-specific motif information. The input prior network must be provided as a three-column, tab-delimited file, where the first column specifies the regulator, the second column the target gene, and the third column the regulatory interaction score. In this study, we used a published [6] prior regulatory network file (`prior.txt`) comprising 4,435,063 edges, constructed from bulk ATAC-seq measurements combined with sequence-specific motif information.

```
9430076C15Rik	1110020A21Rik	0.945914
9430076C15Rik	1110032A03Rik	0.945914
9430076C15Rik	1200011I18Rik	0.945914
9430076C15Rik	1300002E11Rik	0.945914
9430076C15Rik	1500002F19Rik	0.945914
…
Zscan4f	Mcm6	0.0134771
Zscan4f	2610002I17Rik	0.0107817
Zscan4f	1700058P15Rik	0.00539084
Zscan4f	Vasp	0.00539084
Zscan4f	Cgn	0.00269542
```

## *4.3. Applying EstimateNCA to obtain the estimated TFA profiles*

### 4.3.1 Usage of EstimateNCA

We ran EstimateNCA with four λ (regularization parameter) settings, λ=0.000 (unregularized), λ=0.005, λ=0.020 and λ=0.100. The following commands assume that the example input files (`expression.txt`, `regulators.txt`, `targets.txt`, and `prior.txt`) have been downloaded from the MERLIN-SUITE GitHub repository (https://github.com/Roy-lab/MERLIN-

SUITE) and are present in the current working directory. The usages of EstimateNCA run using the data is as below assuming it is installed in `EstimateNCA/NCALearner`:

```
#Unregularized NCA run (λ=0.000) with Rand_init=0
EstimateNCA/NCALearner  -d  expression.txt  -r  regulators.txt  -g
targets.txt           -p      prior.txt      -l      0.000      -o
results/Nca/Lambda_0000/RandInits/Rand_init_0
#Regularized NCA run (λ=0.005) with Rand_init=25
EstimateNCA/NCALearner  -d  expression.txt  -r  regulators.txt  -g
targets.txt           -p      prior.txt      -l      0.005      -o
results/Nca/Lambda_0005/RandInits/Rand_init_25
#Regularized NCA run (λ=0.020) with Rand_init=40
EstimateNCA/NCALearner  -d  expression.txt  -r  regulators.txt  -g
targets.txt           -p      prior.txt      -l      0.020      -o
results/Nca/Lambda_0020/RandInits/Rand_init_40
#Regularized NCA run (λ=0.100) with Rand_init=99
EstimateNCA/NCALearner  -d  expression.txt  -r  regulators.txt  -g
targets.txt           -p      prior.txt      -l      0.100      -o
results/Nca/Lambda_0100/RandInits/Rand_init_99
```

The description of each argument in the EstimateNCA run is as follows:

- ***-d*** expression file (tab-separated) with no header (no cell metadata), rows for each gene. In the present study, the `expression.txt` file consists of a 2,100-gene by 4,633-cell matrix (without cell metadata). (see **Section 4.2.1**).
- ***-r*** list of the regulators to be used for a given target. Here we have used 2,683 reported regulators for mouse. (see **Section 4.2.2**).
- ***-g*** list of the target genes. Same row number (i.e., number of genes) as expression file. Here we have used 2,100 genes (see **Section 4.2.3**).
- ***-p*** prior network file (tab-separated). Here we have used 4,435,063 prior regulatory edges informed by bulk ATAC-seq measurements combined with sequence-specific motif information (see **Section 4.2.4**).
- ***-l*** the regularization parameter λ (lambda) controls model regularization. When λ = 0.000, unregularized NCA is applied; for positive λ values (e.g., 0.005, 0.020, 0.100), regularized NCA (NCA-LASSO) is used.
- ***-o*** specifies the output folder for storing EstimateNCA results for each of the 100 random initializations (`Rand_init = 0-99`) (see **Section 4.2**).

### 4.3.2 Outputs of EstimateNCA

EstimateNCA produces two output files: `adj.txt` and `tfa.txt`. The `adj.txt` file contains the updated regulatory edges with regularized regression coefficients, while `tfa.txt` contains the inferred latent TFA profiles for a subset of regulators. In this study, 131 TFA profiles were identified and saved in `tfa.txt`. The `tfa.txt` (a matrix dimension of 131 TFA by 4633 cells) file is as follows:

```
Alx1  0.570941   -0.83527 …
Ar  0.837172     0.159607 …
Arnt  -0.17097   -1.97831 …
Arx -0.13052     0.894101 …
Atf1  -0.569797  -0.793062 …
…
Zfp110     0.126722   0.201939 …
Zfp143     -0.181769  -0.255976 …
Zfp161     -0.416264  -0.529983 …
Zfx -0.580297    -0.35849 …
Zic3  0.55224    0.356394 …
```

Next, the TFA profiles of the 131 regulators were averaged across 100 random initializations and appended with the suffix "_nca" to distinguish TFA profiles from gene expression profiles of the same regulators. Notably, the "_nca" suffix can alternatively be replaced with "_TFA" if desired.

The resulting `tfa_avg_0_99.txt` (a matrix dimension of 131 TFA by 4633 cells) file looks like following:

```
Alx1_nca   0.673105   -0.38819525 …
Ar_nca     0.13416709 -0.55654479 …
Arnt_nca   -0.213648992     -0.98972757 …
Arx_nca    -0.135140926     0.81517808 …
Atf1_nca   0.075423257      -0.38159042 …
…
Zfp110_nca 0.12169172 0.207829307 …
Zfp143_nca -0.298014921     0.12848773261 …
Zfp161_nca -0.25193142      -0.11648612 …
Zfx_nca    -0.3895989 -0.380896417 …
Zic3_nca   0.50783874 -0.148036293 …
```

These mean TFA profiles are subsequently appended to the expression matrix along with the list of regulators.

## *4.4. Preparing input files for MERLIN-P-TFA*

### 4.4.1. Expression file

For MERLIN-P-TFA analysis, we first appended the original expression matrix (`expression.txt`; 2,100 genes) used by the EstimateNCA application (Section **4.2.1**) with the estimated mean latent TFA profiles of 131 regulators (Section **4.3.2**). This resulted in a combined expression matrix of 2,231 genes (2,100 + 131) by 4,633 cells, without cell metadata, generated separately for each λ (lambda) setting. The combined gene-by-cell matrix (`net1_expression_gene_by_cell.txt`) was used as input to the MERLIN-P application for four different λ values. An example of the merged gene-by-cell expression matrix (`net1_expression.txt`) for λ = 0.100 is shown below.

```
Sept11      2.184866   3.061474 …
Sep15 2.983654   4.238418 …
Marc2 1.077569   1.325657 …
Sept7 2.711823   2.964259  …
Aars  2.094418   2.209428 …
…
Zfp110_nca 0.12169172 0.207829307 …
Zfp143_nca -0.298014921     0.12848773261 …
Zfp161_nca -0.25193142      -0.11648612 …
Zfx_nca    -0.3895989 -0.380896417 …
Zic3_nca   0.50783874 -0.148036293 …
```

MERLIN-P considers the input to be in the cell by gene format without cell meta information, therefore, the final version of input expression matrix (`net1_expression.txt`) looks like as follows:

```
Sept11      Sep15 Marc2 …
2.184866   2.983654   1.077569 …
3.061474   4.238418   1.325657 …
2.237097   3.770023   0.894839 …
1.874197   3.214221   0.909119 …
…
0.000000   0.000000   0.000000 …
2.087869   0.000000   0.000000 …
2.965649   0.000000   0.000000 …
0.000000   2.157659   0.000000 …
0.000000   2.188965   0.000000 …
```

### *4.4.2. Initial cluster assignment file*

MERLIN-P requires an initial cluster (module) assignment file (`clusterassign.txt`) corresponding to the genes in the original expression matrix (`expression.txt`; 2,100 genes), which serves as the starting point for iterative reassignment and refinement of gene module memberships until convergence (see **Theory section**). The input cluster assignment file is formatted as shown below.

```
Sept11      1
Sep15 2
Marc2 3
Sept7 4
Aars  5
…
Zic3  2096
Zscan10    2097
Zwilch     2098
Zwint 2099
Zyx   2100
```

### *4.4.3. List of regulators file*

The regulator list file (`regulators.txt`) generated for the EstimateNCA application (Section **4.2.1**) was appended with the 131 TFA genes (Section **4.3.2**) to produce the regulator list (`net1_transcription_factors.tsv`) used as input for MERLIN-P, as shown below.

```
0610010K14Rik
100041979
1700024P04Rik
1700054O13Rik
1810007M14Rik
…
Zfp110_nca
Zfp143_nca
Zfp161_nca
Zfx_nca
Zic3_nca
```

#### 4.4.4. *Prior Network file*

Because TFA profiles were incorporated into both the input expression matrix (Section **4.4.1**) and the regulator list (Section **4.4.3**), a corresponding update was also applied to the prior network file. The original prior network file (`prior.txt`), used for the EstimateNCA application (Section **4.2.4**), consists of TF-target interactions with three columns: (i) TF regulator name, (ii) target gene name, and (iii) normalized regulatory strength ranging from 0 to 1. This network comprises 4,435,063 TF-target edges from a published motif-based prior network of mouse regulators [6].

To incorporate TFA regulators, each regulatory interaction was duplicated by appending the suffix "_nca" to the corresponding TF regulator names. This expansion resulted in a final prior network file (`net1_net.txt`) containing 8,870,126 edges, which was used as input for the MERLIN-P run.

```
9430076C15Rik	1110020A21Rik	0.945914
9430076C15Rik_nca	1110020A21Rik	0.945914
9430076C15Rik	1110032A03Rik	0.945914
9430076C15Rik_nca	1110032A03Rik	0.945914
…
Zscan4f	Vasp	0.00539084
Zscan4f_nca	Vasp	0.00539084
Zscan4f	Cgn	0.00269542
Zscan4f_nca	Cgn	0.00269542
```

#### 4.4.5. *Configuration file*

The MERLIN-P configuration file (`net1_config.txt`) is a three-column, tab-delimited file in which each row corresponds to a prior network. The first column specifies the network name, the second column provides the file path to the prior network, and the third column indicates the network confidence, where higher values confer greater influence of the prior during model inference.

```
net1	net1_net.txt	5
```

#### 4.4.6. *Subsample files*

To reduce computational burden and enable consensus confidence-based GRN inference, we subsampled the expression matrix (`net1_expression.txt`; **Section 4.4.1**) by randomly partitioning the full dataset (4,633 cells) into half-sized (50%) subsets of 2,317 cells. This subsampling procedure was repeated 50 times using independent random partitions. Each subsample directory (`Subsamples_n2317`) contains 50 index files (`dataindices0.txt-dataindices49.txt`) specifying the selected cells, along with the corresponding subsampled expression matrices of 2,231 genes and TFAs (`dataset0.txt-dataset49.txt`). A summary of all subsampled datasets is provided in `subsample_n2317_list.txt`.

The `Subsamples_n2317/dataset0.txt` file (total lines 2,317) is as follows:

```
1
4
5
…
4630
4631
4632
```

The file `Subsamples_n2317/dataset0.txt` contains expression data for 2,317 cells, formatted with 2,318 rows (including a header row for gene names) and 2,231 columns.

```
1110038B12Rik    1500012F01Rik    1500015O10Rik
1.59324     0.988086    0
1.64028     1.01726     0
…
0     2.09703     0
0     0     0
0     0     0
```

The `subsample_n2317_list.txt` file lists the 50 dataset files:

```
dataset0
dataset1
…
dataset48
dataset49
```

This subsampling step is optional and is included to reduce computational load. Users may alternatively use the original, full expression matrix directly, as implemented in the official MERLIN-P workflow (https://github.com/Roy-lab/merlin-p).

### *4.5. Inferring GRN with MERLIN-P using expression and estimated TFA profiles*

We ran MERLIN-P on 20 subsamples from mouse reprogramming dataset (**Figure 1**) for each λ (lambda) setting using the inputs described in **Section 4.4**. In practice, we recommend running MERLIN-P with 50-100 subsamples. An example command for running MERLIN-P with the example dataset is shown below, assuming the software is installed in `merlin-p/merlin`:

```
#Subsample-based MERLIN-P implementation
merlin-p/merlin -d Subsamples_n2317/dataset0.txt -l
net1_transcription_factors.tsv -q net1_config.txt -c
clusterassign.txt -o out.0/
#Original MERLIN-P implementation
./merlin -d net1_expression.txt -l net1_transcription_factors.tsv -
q net1_config.txt -c clusterassign.txt -o out_dir/
```

#### *4.5.1. Usage of MERLIN-P*

The description of each argument in the MERLIN-P run is as follows:

- ***-d*** expression file, original or subsampled (see **Section 4.4.1 and Section 4.4.6**).
- ***-l*** list of regulators including TFA regulators (see **Section 4.4.3**).

- ***-q*** configuration file containing prior network information (see **Section 4.4.5**).
- ***-c*** initial cluster (module) assignment file (see **Section 4.4.2)**.
- ***-o*** output folder location.

### *4.5.2. Outputs of MERLIN-P*

MERLIN-P outputs regulatory edges between regulators and target genes, which are available in the result folder. The output folder includes a subfolder named `fold0`. Inside the fold0 folder, there are five files: `iter.txt`, `last.txt`, `modules.txt`, `pll.txt`, and the most important file is `prediction_k300.txt`, which contains the inferred regulatory network. The format of the regulatory network file is similar to the input prior network, with the first column specifying the regulator, the second column the target gene, and the third column represents the regression coefficient. Example lines from the inferred regulatory network file `prediction_k300.txt` are as follows:

```
Atf2_nca	1110038B12Rik	0.123109
Id3	1110038B12Rik	0.0377104
```

Another important output file, `modules.txt`, contains the final module assignments of all genes.

```
1500015O10Rik	0
1810058I24Rik	1
2310022B05Rik	2
2410015M20Rik	3
2810004N23Rik	4
9530068E07Rik	5
Aatf	6
…
Tk1	1132
Top2a	1132
Tpx2	1132
Ube2c	1132
Klf6	1133
Vasp	1133
```

## ***4.6. Evaluating and validating the results***

### *4.6.1. Generating high confidence consensus gene regulatory networks*

Since MERLIN-P was run on 20 subsamples, we generated consensus networks across all subsamples and filtered edges using an 80% confidence threshold, retaining only those edges that appeared in at least 16 out of 20 subsamples (**Figure 2a**).

### *4.6.2. AUPR and F-score evaluation using mouse embryonic stem cell (mESCs) gold-standard networks*

We assessed the global accuracy of the inferred consensus networks (confidence≥0.8) across $\lambda$ settings ($\lambda$=0,0.005,0.02,0.1) using the Area Under the Precision-Recall curve (AUPR) and F-score metrics. AUPR quantifies overall network reconstruction performance by summarizing precision-recall trade-offs relative to a gold-standard regulatory network. The F-score is computed as the harmonic mean of precision and recall based on the overlap between inferred edges and gold-standard network edges.

For this evaluation, we used established mouse embryonic stem cell (mESC) gold-standard regulatory networks derived from RNA-seq and microarray data. These gold standards integrate multiple sources of regulatory evidence, including perturbation or knockdown (KD), ChIP-seq (Chip) and combined Chip+KD datasets (see the **README** of https://github.com/Roy-lab/MERLIN-SUITE/), as described in previous studies [6, 9]. Across datasets, we observed highly comparable performance across λ values, with λ=0.1 showing a slight improvement in the both AUPR and F-score in most settings (**Figure 2b**).

### *4.6.3. Generating co-clustering matrix to get the top functionally enriched modules*

We used the module assignment files from the 20 subsamples to generate a co-clustering matrix for each λ setting, which is a (#target genes x #target genes) matrix. The (i,j)-th entry represents the frequency with which genes i and j were assigned to the same module across the 20 subsamples, for example, a frequency of 0.5 indicates that the two genes were clustered together in a module in 10 out of 20 subsamples. The resulting co-clustering matrix is symmetric with diagonal entries equal to 1. Because only target genes are considered, TF regulators are excluded, yielding a (2,100 x 2,100) matrix.

Consensus module assignments were derived using co-clustering score cutoffs ranging from 0.1 to 0.4 for each λ (lambda) value (**Figure 2c**). Modules with at least five constituent genes (M5) were further evaluated for Gene Ontology (GO) enrichment. The proportion of enriched modules was calculated as the number of modules with ≥5 genes showing GO enrichment (GOM5) divided by the total number of modules with ≥5 genes (M5). For λ = 0.100, both 0.2 and 0.4 cutoffs gave a high proportion of enriched modules; however, a 0.4 cutoff reduced gene coverage, so a cutoff of 0.2 was chosen for downstream analysis to obtain 41 modules (**Figure 2d**). In addition to GO enrichment, regulator enrichment analysis was performed for the 41 modules, though this information was not used to select the hyper-parameter setting for further exploration. The inferred consensus network shows these 41 modules (≥5 genes each) represented with distinct colors, and the top regulator for each of the 21 functionally enriched modules is indicated (**Figure 3a**). The number of regulators enriched per module (**Figure 3b;** https://github.com/Roy-lab/MERLIN-SUITE/blob/main/results/Merlinp/Lambda_0100/regulator_enrichAnalysis_0_2_details.txt) and key (top ranked per module) GO processes are depicted (**Figure 3c;** https://github.com/Roy-lab/MERLIN-SUITE/blob/main/results/Merlinp/Lambda_0100/go_enrichAnalysis_0_2_details.txt).

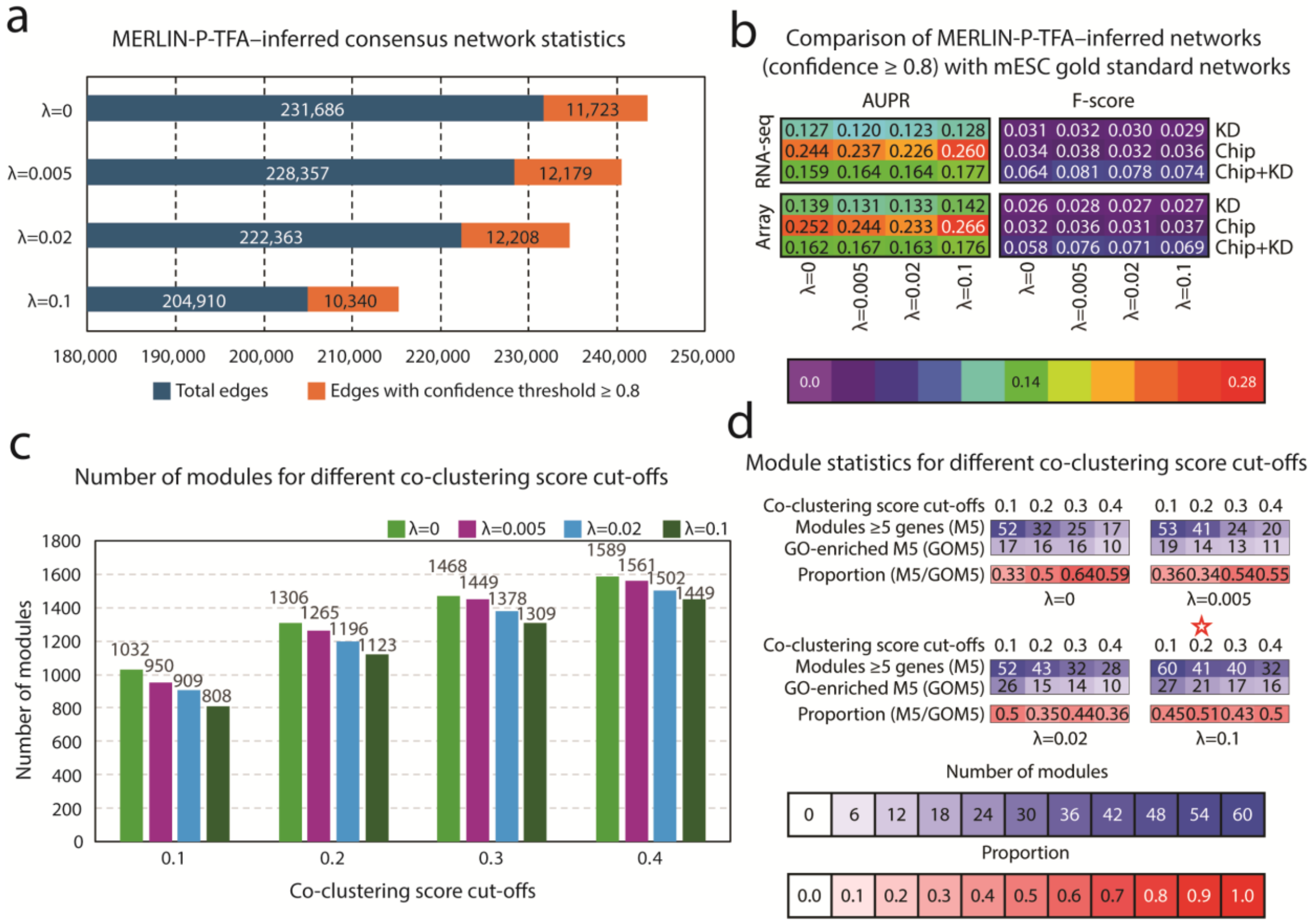

***Figure 2: Evaluation of MERLIN-P-TFA–inferred consensus regulatory networks and module statistics under different regularization parameters (λ) and co-clustering score cut-offs. (a)*** *Total number of inferred regulatory edges and high-confidence edges (confidence threshold ≥ 0.8) generated from RNA-seq datasets using different prior information sources (KD, Chip, and Chip+KD) across varying λ values (0, 0.005, 0.02, and 0.1).* ***(b)*** *Performance comparison of inferred networks against the mESC gold-standard regulatory network using area under the precision–recall curve (AUPR) and F-score metrics for both RNA-seq and array datasets.* ***(c)*** *Number of gene modules identified at different co-clustering score cut-offs (0.1–0.4) for each λ setting.* ***(d)*** *Module quality assessment showing the number of modules containing at least five genes (M5), the number of Gene Ontology–enriched modules (GOM5), and the proportion of GO-enriched modules across different co-clustering score thresholds and λ values. For downstream analyses, we selected the biologically most informative consensus network generated with co-clustering threshold 0.2 and λ = 0.1, which showed the highest proportion of GO-enriched modules (0.51; marked with a red asterisk).*

### *4.7. Approaches to interpret MERLIN-P-TFA output*

The MERLIN-P-TFA inferred network data can be used to identify key regulators, novel regulator-target gene interactions and enriched pathways. Comparisons of module gene expression and network structure can be made across clusters, conditions or timepoints to identify impacts of perturbation or drivers of cellular dynamics. In this study, we demonstrate four approaches for interpreting the MERLIN-P-TFA inferred networks (see the **README** of https://github.com/Roy-lab/MERLIN-SUITE/).

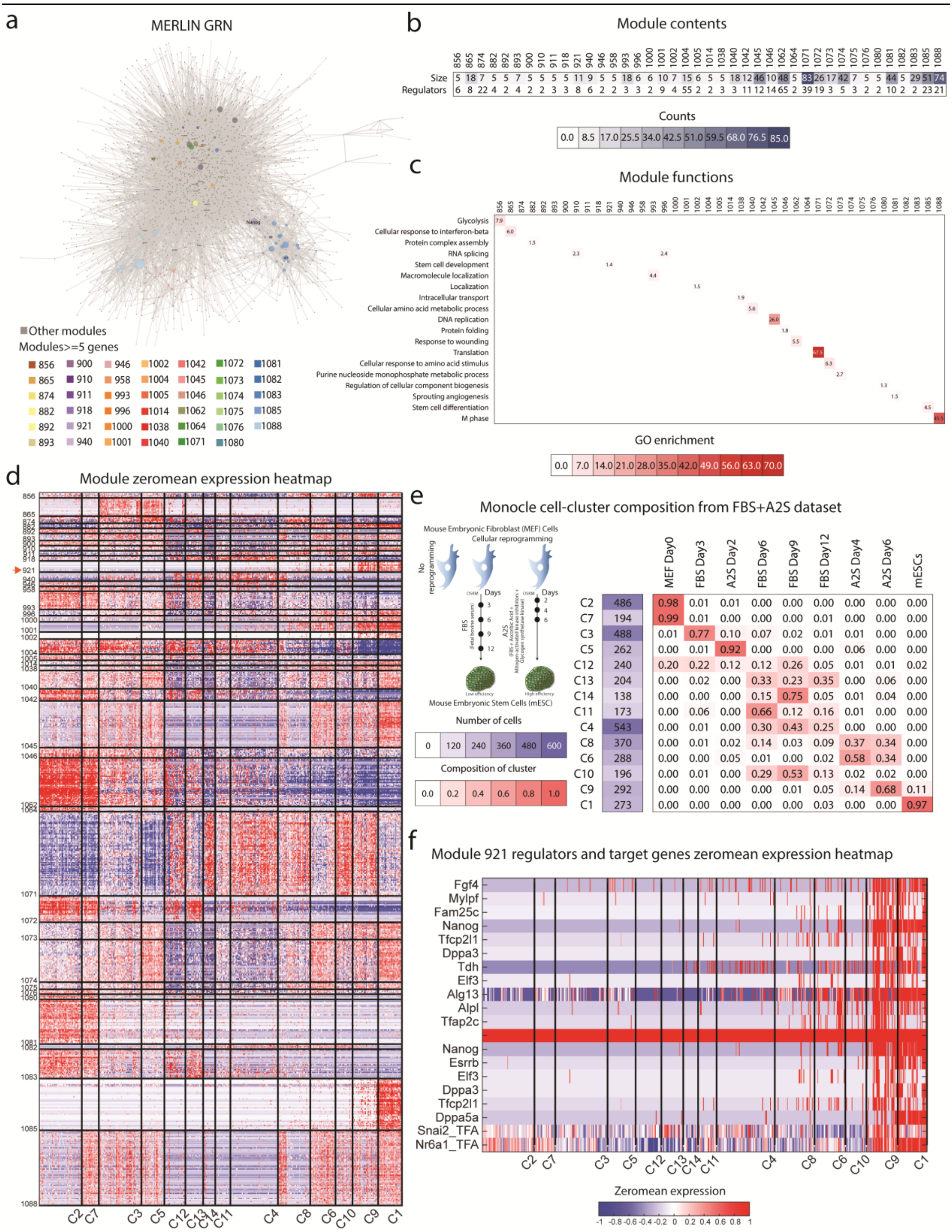

***Figure 3: MERLIN-P-TFA-inferred consensus GRNs and co-expression modules in mouse cellular reprogramming. (a)*** *MERLIN-P-TFA-inferred consensus gene regulatory network (GRN) with a confidence threshold of 80% (edges present in at least 16 out of 20 MERLIN-P-TFA subsample runs) from single-cell RNA-seq of the mouse cellular reprogramming lineage. The network contains 2,100 nodes and 10,340 edges. It comprises 41 co-expression modules (≥5 genes), visualized in distinct colors; genes not assigned to modules are shown in grey ("Other"). Node size is scaled*

*continuously from 5 to 60 based on degree (min-max: 0–61), and node color transparency is adjusted over 200 points. Top regulators (highest regulator enrichment) from functionally (GO) enriched modules (see panel c) are labeled in black Arial font, with size proportional to their outdegree. The network was visualized in Cytoscape using a prefuse force-directed layout. Edges are grey (2-point width) with color transparency adjusted over 120 points.* ***(b–c)*** *Overview of the 41 MERLIN-inferred modules (≥5 genes).* ***(b)*** *shows total genes and regulators per module.* ***(c)*** *shows GO enrichment (top ranked per module) significance (-log10 adjusted p-value). Heatmaps represent counts and GO enrichment confidence.* ***(d)*** *Zero-mean expression heatmap (scaled: lowest:mid:highest = -1:0:1) of all genes present in the 41 modules across cell clusters (C1–C14).* ***(e)*** *Heatmap summary of cell composition in the mouse cellular reprogramming (FBS + A2S) dataset. Cell clusters (C1–C14, y-axis) are shown across samples (x-axis). Total cell counts and proportions are indicated using white–blue (lowest–highest) and white–red (lowest–highest) color scales, respectively.* ***(f)*** *Zero-mean expression heatmap of 8 regulators (below to the red line) and their 11 target genes (above to the red line) from Module 921 per cell cluster. Notably, Module 921 was selected for demonstration due to its importance in stem cell development (see panel c) and its high enrichment in mESC cells (see panel d).*

### *4.7.1. Zeromean expression profile of MERLIN-P-TFA inferred modules*

The zeromean expression profile of the 41 modules is displayed as a heatmap, where the x-axis represents cell clusters (C1-C14) from the mouse cellular reprogramming dataset (**Figure 3e**) and the y-axis represents the 41 modules. This visualization helps to demonstrate the covariation pattern of genes. The module heatmap values correspond to the zeromean expression of constituent genes and regulators, including some MERLIN-P-TFA-inferred TFA profiles (**Figure 3d**). We highlighted an interesting module, Module 921 (**Figure 3f**), which is highly active in the C1 cell cluster (predominantly red compared to other clusters) and is strongly enriched (97%) in mouse embryonic stem cells (mESCs) (**Figure 3e**). Module 921 comprises 11 target genes (*Fgf4, Mylpf, Fam25c, Nanog, Tfcp2l1, Dppa3, Tdh, Elf3, Alg13, Alpl,* and *Tfap2c*), predicted to be regulated by eight key regulators, including six transcription factors (*Nanog, Esrrb, Elf3, Dppa3, Tfcp2l1,* and *Dppa5a*) and two MERLIN-inferred latent TFAs (*Snai2* and *Nr6a1*). Notably, Nanog, Elf3, Dppa3, and Tfcp2l1 appear as both regulators and targets, indicating reciprocal regulatory interactions in which these genes both regulate downstream targets and are themselves regulated by other inferred factors (**Figure 3f**). Importantly, the MERLIN-P-TFA framework captures not only canonical TF regulators based on expression patterns, but also latent transcription factor activities inferred from regulatory structure, which are not directly observable from mRNA abundance alone. This ability to recover hidden regulatory signals highlights a key strength of the MERLIN-P-TFA approach compared to existing network inference methods.

### *4.7.2 Cell-cluster- and sample/condition-specific GRNs visualization and interpretation using MERLIN-VIZ*

MERLIN-VIZ is an R Shiny-based application (https://github.com/Roy-lab/MERLIN-VIZ) for visualizing MERLIN-inferred gene regulatory networks (GRNs), supporting analyses that consider all cells jointly or cells from individual cell-clusters or samples/conditions, and is applicable to both bulk and single-cell-derived GRNs. The application can be downloaded directly from the `MERLIN-VIZ` GitHub repository (https://github.com/Roy-lab/MERLIN-SUITE/tree/main/visualization/MERLIN-VIZ_module_network_visualization) or from the `merlin_viz_singlecell` branch of the MERLIN-VIZ GitHub repository (https://github.com/Roy-lab/MERLIN-VIZ/tree/merlin_viz_singlecell). To initialize the application, the script `net_data_setup.R` must be run once together with `aux_functions.R` to generate the input files `net_data_cluster.Rdata` (for cell-cluster–specific analyses) and `net_data_sample.Rdata` (for sample/condition–specific analyses) file. These files serve as inputs to the corresponding MERLIN-VIZ Shiny applications, `app_cluster.R` and `app_sample.R`, respectively. The `net_data_setup.R` script needs several input files from MERLIN outputs: (i) the inferred network structure (edge list; see **Section 4.5.2**), (ii) node annotations (gene list from the inferred network), (iii) module assignments (module-to-gene mappings and module summaries; see **Section 4.5.2**), and (iv) gene expression data (a gene by cell matrix with gene and cell identifiers). The user can also include optional annotation files, such as gene descriptions, Gene Ontology enrichment results, or regulator enrichment results, to improve network interpretation and visualization. After launching `app_cluster.R` or `app_sample.R`, the application can be accessed locally through a web browser at: `http://127.0.0.1<local port #>` (e.g., [http://127.0.0.1:8888](http://127.0.0.1:8888)).

The `MERLIN-VIZ` or `merlin_viz_singlecell` application (**Figure 4a**) enables visualization of MERLIN-inferred gene regulatory networks (GRNs) across cell clusters (C1-C14) (**Figure 4c**) in the mouse cellular reprogramming dataset, by highlighting context-specific variation in the inferred edge weights (i.e., conditional correlation structure) within a shared network topology (**Figure 4b**). In addition to cell-cluster-specific analyses, MERLIN-VIZ can automatically generate analogous sample- or condition-specific network visualizations (**Supplementary Figure 1**).

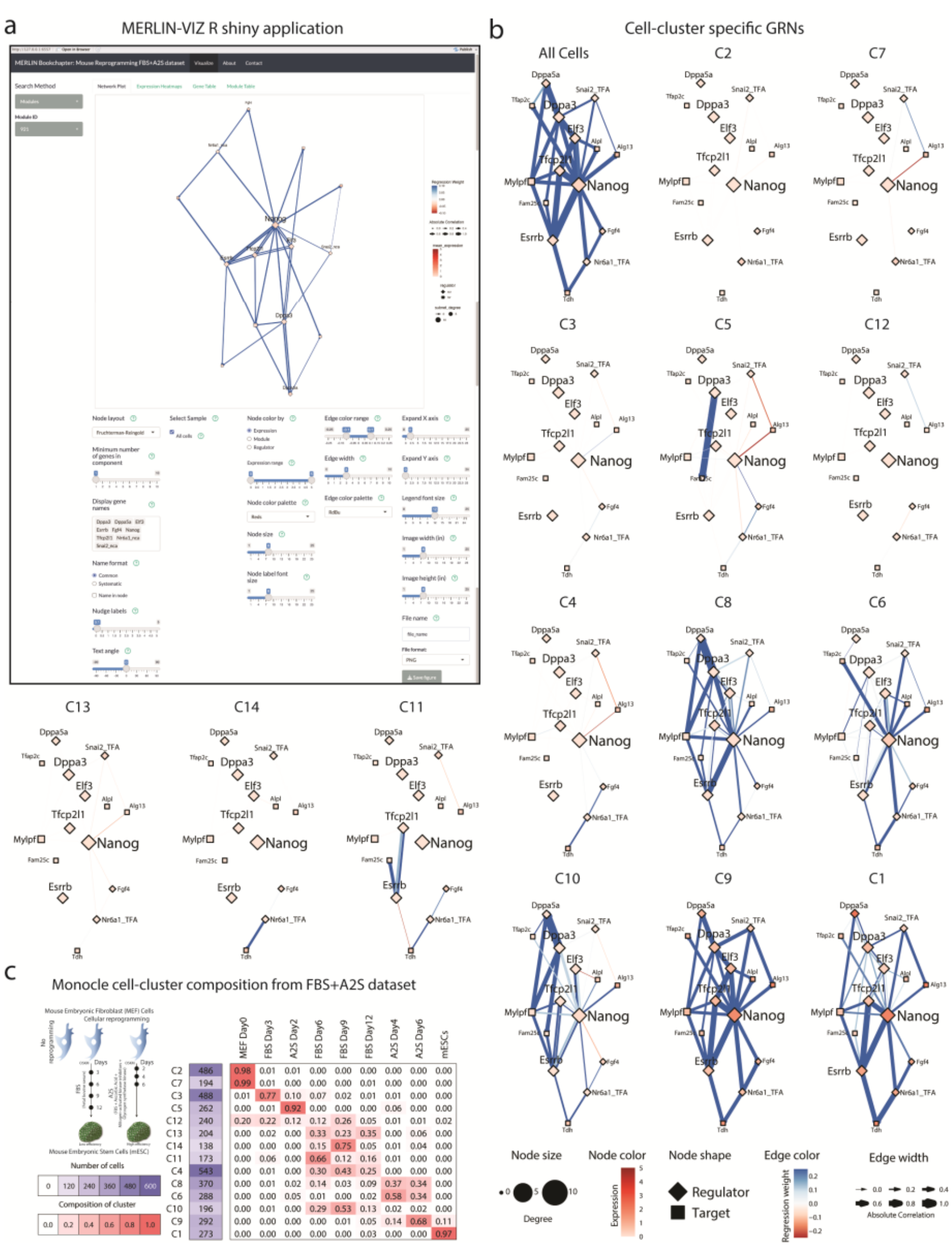

*Figure 4: **Cell-cluster-specific dynamic regulatory network rewiring of Module 921 during reprogramming visualized using the MERLIN-VIZ R Shiny application. (a)** Overview of the*

*MERLIN-VIZ R Shiny application interface (available at https://github.com/Roy-lab/MERLIN-VIZ), which enables interactive visualization of MERLIN-inferred gene regulatory networks (GRNs) at both global and cell-cluster–specific levels. Users can customize node, edge, and network attributes and export high-resolution figures.* ***(b)*** *Cell-cluster-specific dynamic GRNs inferred by MERLIN-P-TFA. Nodes represent regulators (diamonds) or target genes (squares), with node size indicating network degree and node color representing mean expression. Edges denote regulatory interactions, with width proportional to absolute correlation and color indicating regression weight (activation or repression). This depiction indicates the Nanog related change of regulatory interaction dynamics over different cell cluster and the connectivity was observed highest in C1 cell clusters which is correspond the mESC cells (see panel c).* ***(c)*** *Pictorial depiction of the experimental design of the mouse cellular reprogramming dataset (left), summary of the cell composition of different cell-clusters (C1-C14) across nine experimental conditions (MEF Day 0, FBS time points, A2S time points, and mESCs; right).*

Using the cell-cluster-specific visualization framework, we identified hub regulators and regulator-specific network rewiring across distinct cell clusters. For example, Dppa3 exhibits prominent rewiring in the C5 cluster (**Figure 4b**), which is composed predominantly of A2S Day 2 cells (92%) (**Figure 4c**). This early-to-mid stage activity of Dppa3 (Developmental Pluripotency Associated 3; Stella) during MEF-to-mESC reprogramming, including its predicted regulation of Fam25c (Family With Sequence Similarity 25 Member C) (**Figure 4b**), is consistent with its known role in epigenetic reprogramming and germline gene activation, supporting a transitional regulatory state bridging early reprogramming and later pluripotency circuits [14]. In contrast, Esrrb (Estrogen Related Receptor Beta) shows extensive regulatory rewiring across clusters C11, C8, C6, C9, and C1 (**Figure 4b**), corresponding to later stages of FBS (Days 6, 9, and 12) and A2S (Days 4 and 6) reprogramming toward the mESC lineage (**Figure 4c**), suggesting a critical role in late-stage fate determination. Esrrb acts downstream of LIF–Jak/Stat3 signaling to promote the acquisition of naive pluripotency and has been shown to sustain pluripotent identity through mechanisms that are at least partially independent of Nanog [15]. A similar pattern is observed for Nanog (Nanog Homeobox), whose regulatory rewiring is most pronounced in clusters C8, C6, C10, C9, and C1 (**Figure 4b**), which are dominated by cells from later A2S stages (Days 4 and 6) and the mESC endpoint (**Figure 4c**). This observation is consistent with previous studies highlighting the essential role of Nanog in establishing and stabilizing pluripotency during cellular reprogramming [11].

#### *4.7.3 Condition-specific GRNs visualization and interpretation using Cytoscape*

In addition to cell cluster– and sample/condition–specific GRN visualization provide by MERLIN-VIZ (**Section 4.7.2**; **Figure 4** and **Supplementary Figure 1**), MERLIN-inferred networks can also be exported and visualized in Cytoscape, offering an alternate platform for exploring condition-specific regulatory rewiring. The workflow for generating and visualizing condition-specific GRNs in Cytoscape is described in detail in the MERLIN-SUITE GitHub repository: https://github.com/Roy-lab/MERLIN-SUITE/blob/main/README.md. To be consistent with **Figure 4**, we focused on Module 921 for condition-specific visualization across stages of mouse cellular reprogramming (**Figure 5**). Note that MERLIN-VIZ also offers network visualizations per sample (**Supplementary Figure 1**) but we used Cytoscape as an alternative giving users more manual control over graph layout and attributes. MERLIN-VIZ offers control for updating node and edge layouts, but we show the default, automated visualization without any fine-tuning to improve the visualization.

In Module 921, Nanog emerges as a prime hub regulator (node degree: 13), with its expression gradually increasing over the course of MEF-to-mESC reprogramming and dynamically modulating its regulatory interactions with other factors (**Figure 5b**). Nanog, together with the canonical OSKM factors (Oct4, Sox2, Klf4, c-Myc), is essential for establishing full pluripotency and maintaining the stability of pluripotent and germline-like states [16]. In this study, the Nanog-Tfcp2l1 (Transcription Factor CP2 Like 1) regulatory relationship exhibits a temporal switch in interaction polarity. Tfcp2l1 is a direct downstream target of the LIF/Stat3 pathway and contributes to the maintenance of naive pluripotency by reinforcing expression of key ESC genes; it can compensate for Nanog under certain conditions and sustain pluripotency even when Nanog is absent [17]. In early MEF cells, both regulators mutually repress each other (blue edges), a pattern maintained during the initial stages of FBS (Days 3 and 6) and A2S (Days 2 and 4) reprogramming. During intermediate stages, such as FBS Day 9, Nanog continues to repress Tfcp2l1, while Tfcp2l1 shifts toward activating Nanog (red edge). At later time points corresponding to fully reprogrammed mESCs in both the low-efficiency FBS and accelerated A2S pathways, Nanog and Tfcp2l1 engage in strong reciprocal activation, illustrating a dynamic rewiring that stabilizes pluripotency during the transition from somatic to embryonic stem cell fate (**Figure 5b**).

Beyond this feedback loop, Tfcp2l1 also activates Esrrb at later reprogramming stages (FBS Day 12, A2S Day 6, mESCs), consistent with previous studies showing direct binding of Tfcp2l1 to the Esrrb promoter, reinforcing naive pluripotency and supporting mESC identity [17]. Additionally, Elf3 (E74 Like ETS Transcription Factor 3) participates in regulatory rewiring, with interactions between Nanog-Elf3 and Tfcp2l1-Elf3 shifting from repressive (blue) to activating (red) connectivity at later stages (FBS Day 12, A2S Day 6, mESCs). By interfacing with Nanog and Tfcp2l1, Elf3 likely facilitates silencing of fibroblast-specific transcriptional programs by mesenchymal-epithelial transition and promotes chromatin accessibility necessary for late-stage pluripotency factor engagement [18]. Its activity appears most pronounced in early-to-intermediate clusters, where somatic identity must be dismantled before full activation of the Nanog-Tfcp2l1-Esrrb pluripotency axis (**Figure 5b**). Together, these findings suggest that early-stage antagonism between Nanog and Tfcp2l1 resolves into a cooperative regulatory network at late stages, with sequential engagement of Esrrb and Elf3 stabilizing the naive pluripotent transcriptional program [18–20].

In addition to canonical transcriptional regulators, MERLIN-P-TFA uniquely identified two latent transcription factor activities (TFAs), Snai2 and Nr6a1, within Module 921 that directly target core pluripotency genes. Snai2_TFA is predicted to regulate Nanog and Alg13, the latter also being a Nanog target, suggesting coordinated control over pluripotency maintenance and chromatin-associated processes. Similarly, Nr6a1_TFA targets Fgf4 and Tdh, which are also regulated by Nanog and Esrrb, respectively, revealing a layer of hidden regulatory influence that complements the canonical pluripotency network. Functionally, Snai2 and Nr6a1 exhibit activity at early stages of MEF-to-mESC reprogramming, where they likely contribute to repression of fibroblast-specific transcriptional programs and priming of chromatin for pluripotency factor engagement [21, 22]. As reprogramming progresses, core regulators such as Nanog and Esrrb become dominant, consolidating the naive pluripotent state, while the early latent activities of Snai2 and Nr6a1 provide foundational regulatory input that shapes the timing, coordination, and robustness of the late-stage pluripotency network (**Figure 5b**; [19, 21, 22]).

Together, these regulatory interactions in Module 921 likely drive the coordinated activation of regulators, target genes and TFA required for mESC fate determination, highlighting key mechanisms by which fibroblast cells are reprogrammed toward a pluripotent state.

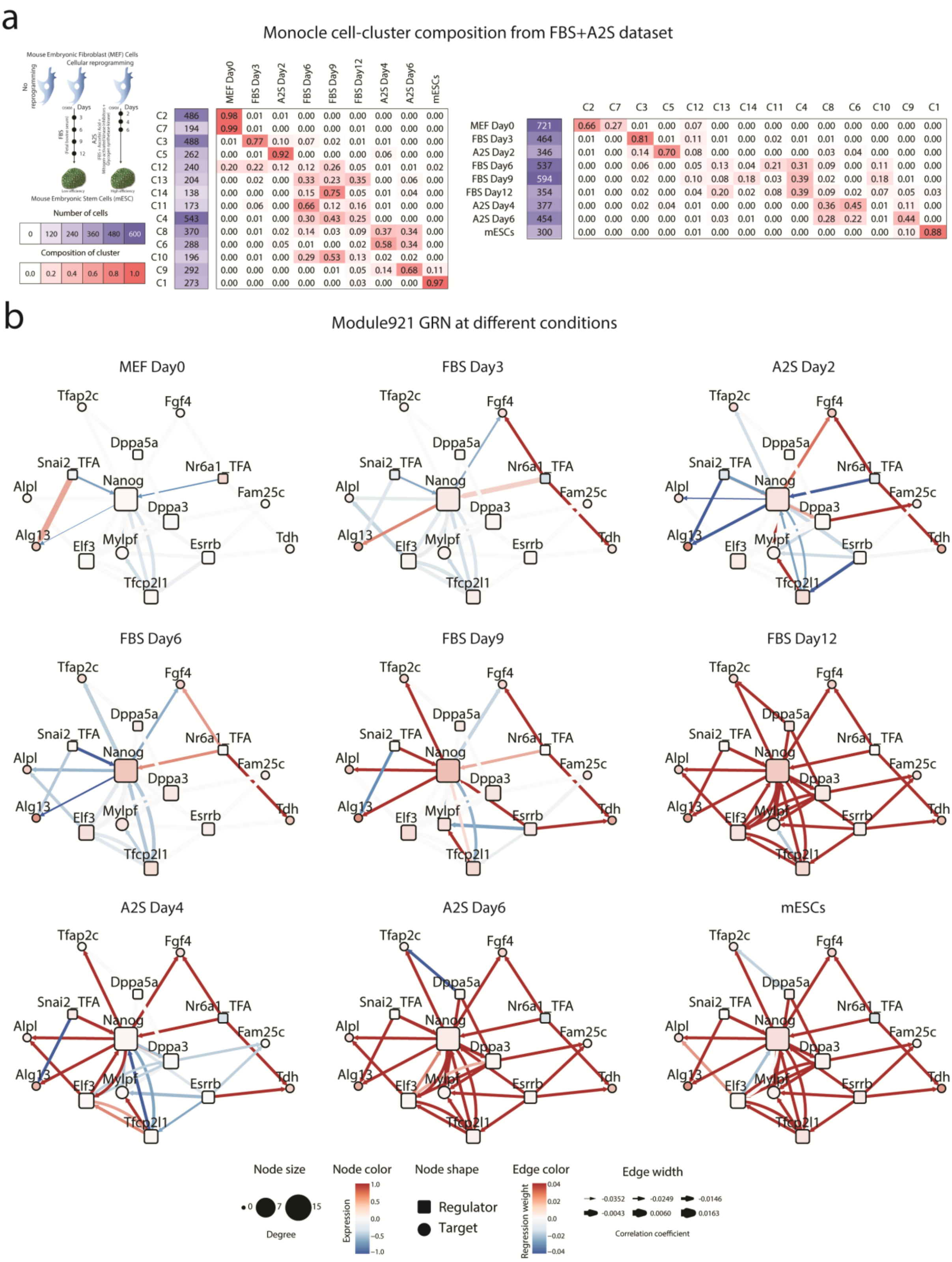

***Figure 5: Condition-specific dynamic regulatory network rewiring of Module 921 during reprogramming. (a)*** *Pictorial depiction of the experimental design of the mouse cellular reprogramming dataset (left), summary of the cell composition of different cell-clusters (C1-C14) across nine experimental conditions (MEF Day 0, FBS time points, A2S time points, and mESCs; middle) and reciprocal representation of cluster contributions to each condition (right).* ***(b)***

*Comparison of MERLIN-P-TFA–inferred gene regulatory network (GRN) structure for Module 921 across the nine conditions. Nodes correspond to regulators (square shape) and target genes (round shapes), with node size reflecting node connectivity and node color indicating expression level. Edge color denotes the signed regression weight inferred by MERLIN-P-TFA, while edge width represents the magnitude of the associated correlation coefficients. Comparison of network structure across conditions reveals a progressive increase in the regulatory influence and connectivity of pluripotency-associated regulators (e.g., Nanog, Dppa3, Dppa5a, Elf3, Esrrb, Tfcp2l1) and selected transcription factor activities (TFAs; e.g., Nr6a1_TFA, Snai2_TFA), accompanied by a reduction in fibroblast-associated regulatory influence. These results demonstrate that MERLIN captures gradual, condition-dependent regulatory reorganization during cellular reprogramming.*

#### *4.7.4 Visualization of functional and cell-cluster specific pseudobulk expression variations of genes in modules*

Another way to visualize MERLIN-inferred results is through pseudobulk expression profiles of modules across cell clusters, integrating the expression of constituent genes (regulators, TFAs, and targets) with their functional annotations and regulatory connectivity (**Figure 6**). Pseudobulk based visualization is helpful for single cell RNA-seq data with a large number of cells. We selected two representative modules, Module 882 (**Figure 6b**) and Module 921 (**Figure 6c**), for illustration based on their high expression across mouse cellular reprogramming cell clusters (**Figure 6a**) and their relevance to earlier analyses, respectively (Sections **4.7.2** and **4.7.3**).

In Module 882, we identified five target genes (Ncl, Hsp90aa1, Npm1, Hsp90ab1, and Trim28) associated with developmental processes, macromolecular protein complex assembly and organization, and regulation of cellular stress and metabolic responses (magenta heatmaps, top-left panel). These genes exhibit consistently high expression across all cell clusters, indicating that Module 882 represents a broadly active regulatory program maintained throughout cellular reprogramming (expression heatmaps, top and bottom right panels). In addition, MERLIN-inferred transcription factor activities (TFAs) of four regulators (Arnt, Hsf4, Pax5, and Max) reveal dynamic rewiring of regulatory control, with pronounced activity at later reprogramming stages, particularly in the C9 and C1 cell clusters (cyan heatmaps, middle panel; **Figure 6b**).

The regulatory features of Module 921 at both the cell cluster–specific (**Figure 4b**; **Section 4.7.2**) and condition-specific (**Figure 5b** and **Supplementary Figure 1**; **Sections 4.7.3** and **4.7.2**) levels have been discussed in detail above. Here, we focus on the functional characterization of genes within this module. Key genes in Module 921, including Nanog, Fgf4, and Tfap2c, are significantly enriched for biological processes related to stem cell development and maintenance (magenta heatmaps, top-left panel). The regulatory connectivity map, integrating regulators, TFAs, and target genes, provides a clear depiction of factor–target relationships, as illustrated in the heatmap visualization (cyan heatmaps, middle panel). Consistent with this regulatory architecture, pseudobulk expression profiles show that Module 921 genes and regulators are predominantly active in the C9 and C1 cell clusters, which are representative of the mESC identity (expression heatmaps, top and bottom right panels; **Figure 6c**).

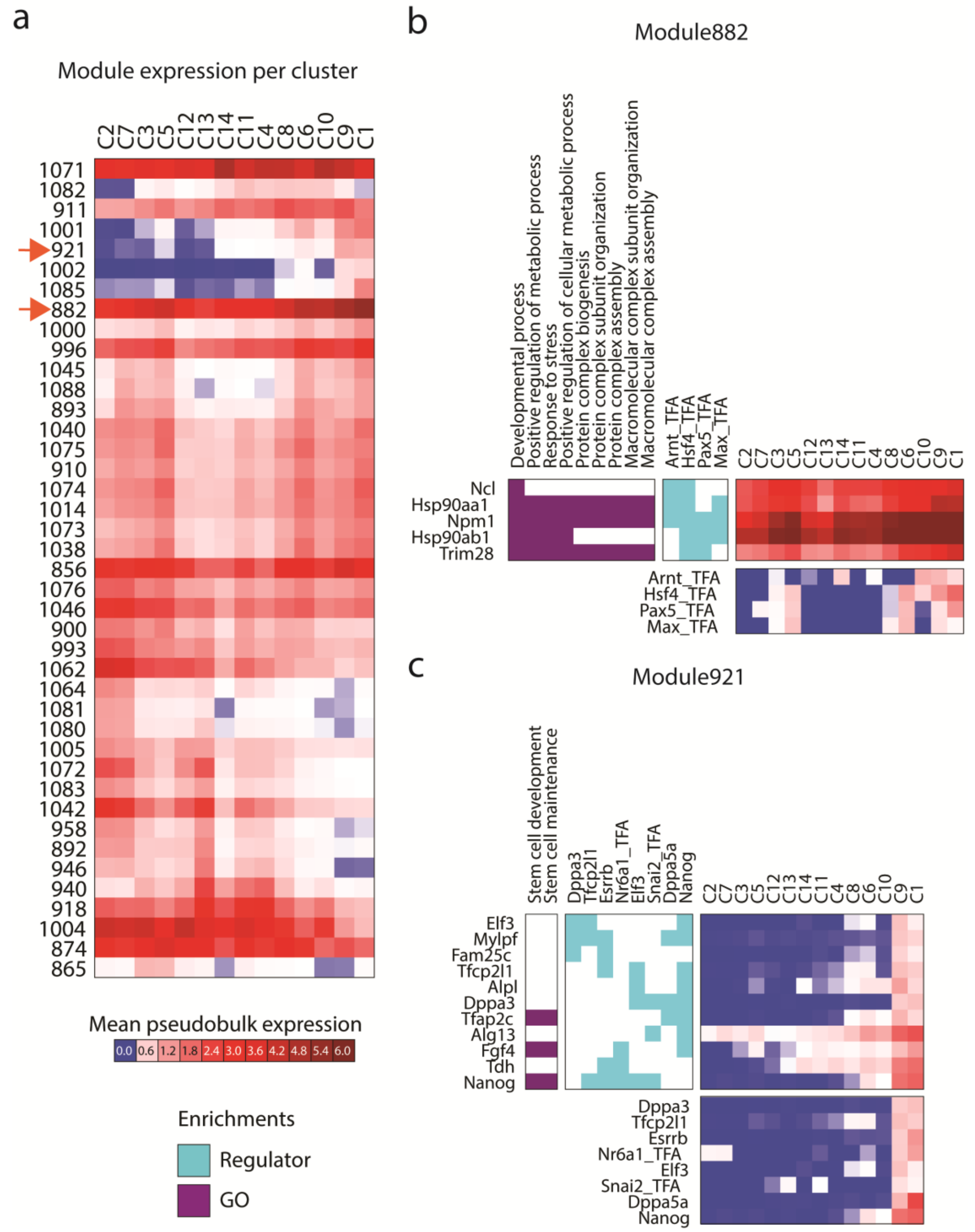

***Figure 6: Functional characterization of the MERLIN-P-TFA-inferred gene regulatory modules.*** ***(a)*** *Mean pseudobulk expression profiles of MERLIN-P-TFA-inferred modules across 14 cell-clusters (C1-C14), illustrating cell-cluster–specific activation patterns.* ***(b-c)*** *Functional annotation of Module 882* ***(b)*** *and Module 921* ***(c)****. Gene Ontology (GO) enrichment analysis was performed on module target genes, and enriched biological processes are shown together with their associated inferred regulators and transcription factor activities (TFAs). Dark magenta indicates enriched GO biological processes linked to target genes, while cyan denotes the corresponding regulators or TFAs associated with those targets. For each module, the top panel shows pseudobulk expression profiles of target genes across cell clusters, and the bottom panel shows pseudobulk expression profiles of the associated regulators and TFAs. Together, these panels connect MERLIN-P-TFA–inferred regulatory structure with functional gene programs active across cell clusters.*

**Acknowledgements**

This work is supported by the National Institutes of Health (Grant ID GM144708-01A1), and the Department of Energy (DOE grants DE-SC0021052, DE-SC0023082). We thank members of the Roy Lab, Spencer Halberg-Spencer, Livvy T. Johnson, Khagani Eynullazada, Dr. Junha Shin, and Sunnie Grace McCalla, for their valuable assistance and discussions related to MERLIN network visualization.

**References**

1. Chasman D, Roy S (2017) Inference of cell type specific regulatory networks on mammalian lineages. Curr Opin Syst Biol 2:130–139. https://doi.org/10.1016/j.coisb.2017.04.001

2. Kim HD, Shay T, O'Shea EK, Regev A (2009) Transcriptional regulatory circuits: predicting numbers from alphabets. Science 325:429–432. https://doi.org/10.1126/science.1171347

3. Thompson D, Regev A, Roy S (2015) Comparative analysis of gene regulatory networks: from network reconstruction to evolution. Annu Rev Cell Dev Biol 31:399–428. https://doi.org/10.1146/annurev-cellbio-100913-012908

4. Vogel C, Marcotte EM (2012) Insights into the regulation of protein abundance from proteomic and transcriptomic analyses. Nat Rev Genet 13:227–232. https://doi.org/10.1038/nrg3185

5. Castro DM, de Veaux NR, Miraldi ER, Bonneau R (2019) Multi-study inference of regulatory networks for more accurate models of gene regulation. PLoS Comput Biol 15:e1006591. https://doi.org/10.1371/journal.pcbi.1006591

6. McCalla SG, Fotuhi Siahpirani A, Li J, Pyne S, Stone M, Periyasamy V, Shin J, Roy S (2023) Identifying strengths and weaknesses of methods for computational network inference from single-cell RNA-seq data. G3 (Bethesda) 13:jkad004. https://doi.org/10.1093/g3journal/jkad004

7. Roy S, Lagree S, Hou Z, Thomson JA, Stewart R, Gasch AP (2013) Integrated module and gene-specific regulatory inference implicates upstream signaling networks. PLoS Comput Biol 9:e1003252. https://doi.org/10.1371/journal.pcbi.1003252

8. Siahpirani AF, Roy S (2017) A prior-based integrative framework for functional transcriptional regulatory network inference. Nucleic Acids Res 45:e21. https://doi.org/10.1093/nar/gkw963

9. Siahpirani AF, McCalla SG, Pyne S, Dillingham C, Sridharan R, Roy S (2025) Uncovering Functional Gene Regulatory Networks in Bulk and Single-Cell Data through Robust Transcription Factor Activity Estimation and Model-Guided Experimental Validation. bioRxiv 2025.06.09.658650. https://doi.org/10.1101/2025.06.09.658650

10. Carriel CC, Halberg-Spencer SA, Kotvanova M, Pyne S, Park SC, Seo H-W, Schmidt A, Calise DG, Ané J-M, Keller NP, Roy S (2026) A network-based model of Aspergillus fumigatus elucidates regulators of development and defensive natural products of an opportunistic pathogen. Nucleic Acids Res 54:gkaf1439. https://doi.org/10.1093/nar/gkaf1439

11. Tran KA, Pietrzak SJ, Zaidan NZ, Siahpirani AF, McCalla SG, Zhou AS, Iyer G, Roy S, Sridharan R (2019) Defining Reprogramming Checkpoints from Single-Cell Analyses of Induced Pluripotency. Cell Rep 27:1726-1741.e5. https://doi.org/10.1016/j.celrep.2019.04.056

12. Heckerman D, Chickering DM, Meek C, Rounthwaite R, Kadie C (2001) Dependency Networks for Inference, Collaborative Filtering, and Data Visualization. J Mach Learn Res 1:49–75. https://doi.org/10.1162/153244301753344614

13. Liao JC, Boscolo R, Yang Y-L, Tran LM, Sabatti C, Roychowdhury VP (2003) Network component analysis: reconstruction of regulatory signals in biological systems. Proc Natl Acad Sci U S A 100:15522–15527. https://doi.org/10.1073/pnas.2136632100

14. Toriyama K, Au Yeung WK, Inoue A, Kurimoto K, Yabuta Y, Saitou M, Nakamura T, Nakano T, Sasaki H (2024) DPPA3 facilitates genome-wide DNA demethylation in mouse primordial germ cells. BMC Genomics 25:344. https://doi.org/10.1186/s12864-024-10192-7

15. Huang D, Wang L, Duan J, Huang C, Tian XC, Zhang M, Tang Y (2018) LIF-activated Jak signaling determines Esrrb expression during late-stage reprogramming. Biol Open 7:bio029264. https://doi.org/10.1242/bio.029264

16. Silva J, Nichols J, Theunissen TW, Guo G, van Oosten AL, Barrandon O, Wray J, Yamanaka S, Chambers I, Smith A (2009) Nanog is the gateway to the pluripotent ground state. Cell 138:722–737. https://doi.org/10.1016/j.cell.2009.07.039

17. Wang X, Wang X, Zhang S, Sun H, Li S, Ding H, You Y, Zhang X, Ye S-D (2019) The transcription factor TFCP2L1 induces expression of distinct target genes and promotes self-renewal of mouse and human embryonic stem cells. J Biol Chem 294:6007–6016. https://doi.org/10.1074/jbc.RA118.006341

18. Subbalakshmi AR, Sahoo S, Manjunatha P, Goyal S, Kasiviswanathan VA, Mahesh Y, Ramu S, McMullen I, Somarelli JA, Jolly MK (2023) The ELF3 transcription factor is associated with an epithelial phenotype and represses epithelial-mesenchymal transition. J Biol Eng 17:17. https://doi.org/10.1186/s13036-023-00333-z

19. Sevilla A, Papatsenko D, Mazloom AR, Xu H, Vasileva A, Unwin RD, LeRoy G, Chen EY, Garrett-Bakelman FE, Lee D-F, Trinite B, Webb RL, Wang Z, Su J, Gingold J, Melnick A, Garcia BA, Whetton AD, MacArthur BD, Ma'ayan A, Lemischka IR (2021) An Esrrb and Nanog Cell Fate Regulatory Module Controlled by Feed Forward Loop Interactions. Front Cell Dev Biol 9:630067. https://doi.org/10.3389/fcell.2021.630067

20. Song S, Heo J, Lee S, Nam YJ, Kim Y, Ju H, Kwon H, Im HJ, Ha SW, Kim HJ, Lee D, Park SJ, Song SH, Park J, Jeong EM, Kim K, Shin D-M, Lee S (2025) Dual Specificity Phosphatase 27 Regulates Pluripotency and Meso-Endodermal Differentiation by Interacting with Transcription Factor CP2 Like-1 in Embryonic Stem Cells. Int J Stem Cells 18:412–425. https://doi.org/10.15283/ijsc25047

21. Gingold JA, Fidalgo M, Guallar D, Lau Z, Sun Z, Zhou H, Faiola F, Huang X, Lee D-F, Waghray A, Schaniel C, Felsenfeld DP, Lemischka IR, Wang J (2014) A genome-wide RNAi screen identifies opposing functions of Snai1 and Snai2 on the Nanog dependency in reprogramming. Mol Cell 56:140–152. https://doi.org/10.1016/j.molcel.2014.08.014

22. Li J, Mascarinas P, McGlinn E (2024) The expanding roles of Nr6a1 in development and evolution. Front Cell Dev Biol 12:1357968. https://doi.org/10.3389/fcell.2024.1357968

***Supplementary Figure 1: Sample/condition-specific dynamic regulatory network rewiring of Module 921 during reprogramming visualized using the MERLIN-VIZ R Shiny application.*** *Comparison of MERLIN-P-TFA-inferred gene regulatory network (GRN) structure for Module 921 across nine experimental conditions (MEF Day0, FBS Days 3, 6, 9, and 12, A2S Days 2, 4, and 6, and mESCs) in the MERLIN-VIZ R Shiny application interface (available at [https://github.com/Roy-lab/MERLIN-VIZ](https://github.com/Roy-lab/MERLIN-VIZ)). Nodes represent regulators (diamonds) or target genes (squares), with node size indicating network degree and node color representing mean expression. Edges denote regulatory interactions, with width proportional to absolute correlation and color indicating regression weight (activation or repression). Comparison of network structure across conditions reveals dynamic rewiring of pluripotency-associated regulators, including Nanog, Dppa3, Esrrb, and Tfcp2l1, during the transition from mouse embryonic fibroblasts (MEFs) to mouse embryonic stem cells (mESCs). The visualization highlights condition-specific changes in regulatory connectivity while preserving a common network topology, thereby facilitating the identification of temporal regulatory programs associated with cellular reprogramming.*

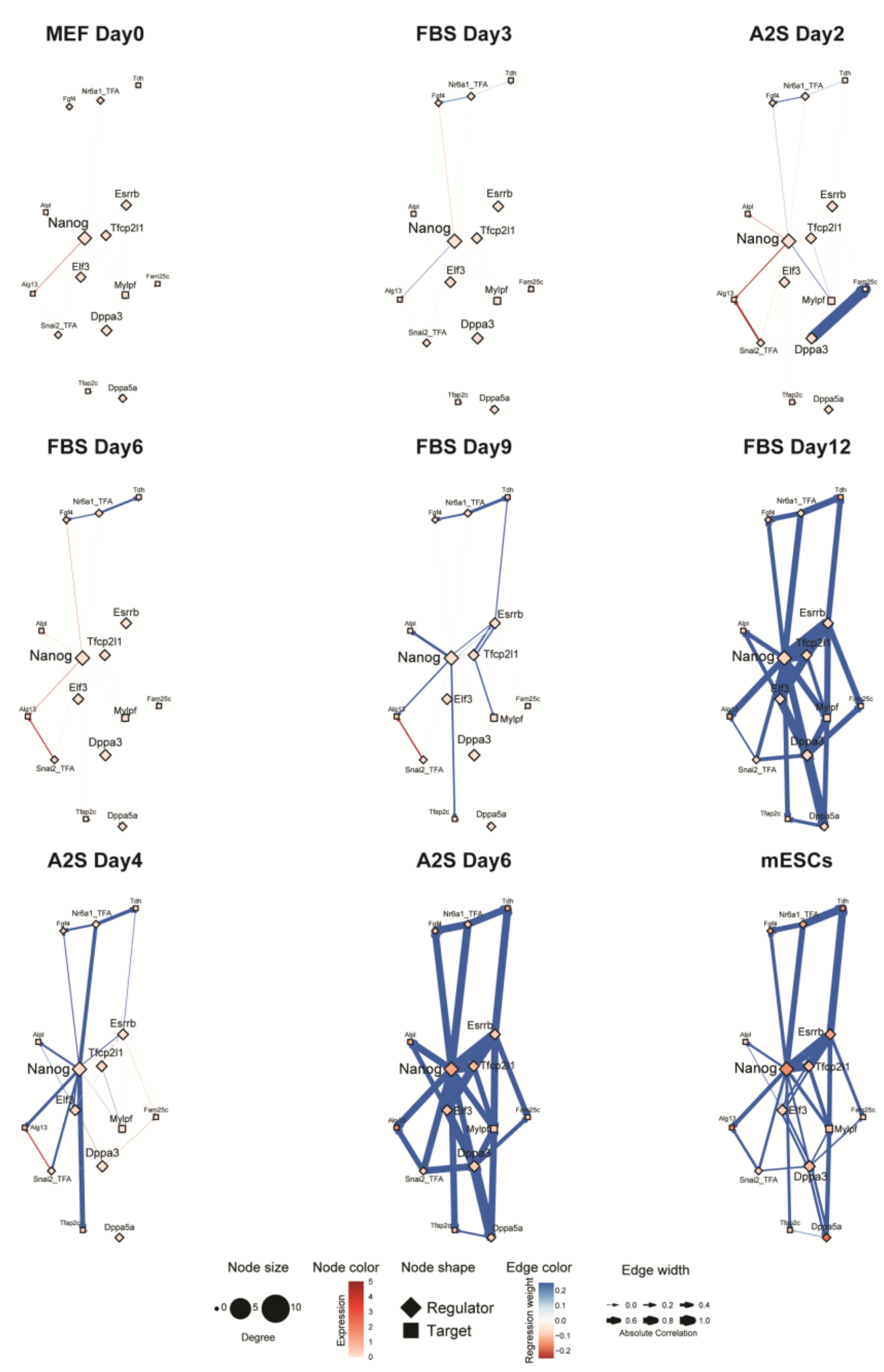